\documentstyle[aps,prl,twocolumn,epsfig]{revtex}
\draft
\begin{document}
\tighten
\onecolumn
\twocolumn[\hsize\textwidth\columnwidth\hsize\csname @twocolumnfalse\endcsname

\title{Global Phase Diagram of a One-Dimensional
Driven Lattice Gas}
\author{Dirk Helbing,$^{1,2}$ David Mukamel,$^2$ and Gunter M.
Sch\"utz$^{2,3}$}
\address{$^1$ II. Institute of Theoretical Physics, Pfaffenwaldring 57/III,
70550 Stuttgart, Germany\\
$^2$ Department of Physics of Complex Systems,
Weizmann Institute of Science, Rehovot, Israel\\
$^3$ Institut f\"ur Festk\"orperforschung, Forschungszentrum J\"ulich,
52425 J\"ulich, Germany}
\maketitle
\begin{abstract}
We investigate the non-equilibrium stationary state of a translationally
invariant one-dimensional driven lattice gas with short-range interactions.
The phase diagram is found to exhibit
a line of continuous transitions from a disordered phase to a
phase with spontaneous symmetry breaking. At the phase transition
the correlation length is infinite and density correlations decay
algebraically. Depending on the parameters which define the dynamics, the
transition either belongs to the universality class of directed
percolation or to a universality class of a growth model
which preserves the local minimal height.
Consequences of some mappings to other models, including a
parity-conserving branching-annihilation process are briefly discussed.
\end{abstract}
]
The interplay of external driving fields and internal repulsive forces
between particles can lead to interesting and unexpected phase transitions in
the steady states of one-dimensional driven diffusive systems even if the
interactions are only short-ranged \cite{Schm96}.
Generically, the presence of boundaries or single defects in
driven systems leads to shock waves
and mutual blocking mechanisms which result in a breakdown of homogeneous
particle flow. Thus localized static inhomogeneities are responsible for a
variety of phenomena including first- and second-order phase transitions
\cite{phase} or spontaneous symmetry breaking
\cite{sym}. These observations are of practical importance for the
qualitative understanding of many-body systems in which the dynamic degrees of
freedom reduce to effectively one dimension as e.g. in traffic flow
\cite{traffic}, kinetics of protein synthesis \cite{protein},
gel-electrophoresis \cite{gel},
or interface growth of thin films \cite{Blak79}.

Whether continuous phase transitions can occur also in spatially
{\em homogeneous} non-equilibrium systems in one dimension is less
well-understood \cite{abs}. In particular, there is a long-stand\-ing
conjecture \cite{Spoh91} that in systems with local interactions the steady
states have rapidly decaying correlations and,
like in 1-d {\em equilibrium} models, no phase transition
accompanied by algebraically decaying correlations takes place.
On the other hand, recent studies of more complicated driven systems
of three or more species of particles in $1d$ have demonstrated that
phase separation may take place in these models \cite{Evans},
thus proving the possibility of long-range order, but leaving open
the issue of continuous phase transitions with algebraic decay of correlations.
In the absence of a general framework for studying non-equilibrium
phase transitions, analyzing specific models could provide useful
insight in these complex phenomena.

In this context, several translationally invariant one-dimensional
growth models with local interactions which exhibit roughening transitions
have recently been introduced. A common feature of these models is that
one of the local transition rates which govern their
dynamics is set to zero. The resulting roughening transition in one class of
models belongs to the universality class of directed percolation
\cite{Alon96}. In another class of growth models which
preserve the local minimal height, the transition is found to belong to a
different universality class \cite{Kodu98,Hinr97}. It would be of great
interest to put
these classes of models within a unifying framework, so that the various
types of transitions, the associated crossover phenomena and the global
phase diagram could be studied.

In this Letter we introduce a simple homogeneous driven $1d$ lattice gas model
with local dynamics. It exhibits a phase transition where correlations decay
algebraically and which is accompanied by spontaneous symmetry
breaking. The model can be mapped onto a growth model where the transition
becomes a roughening transition. By varying the parameters which define its
dynamics, some types of the transitions discussed above can be realized. The
various transitions and the global phase diagram are studied.

We consider a lattice gas which is an asymmetric exclusion
process with next-nearest-neighbour interaction.
Each lattice-site
$i\in\{1,2,\dots,L\}$ of a periodic chain may be either empty
($\emptyset$) or occupied by one particle of a single species,
labeled $A$. The model evolves by random sequential updating.
Particles hop to the right with constant attempt rate $r$ ($q$)
if the right nearest neighbour site is vacant and
the nearest neighbour site at the left is occupied (empty).
Unlike in the KLS-models \cite{Katz84}, the left-hopping mechanism is
different:
A particle hops to the left with rate $p=1-q-r$ only if the
next-nearest-neighbour site is empty as well. The model is therefore
defined by the transitions
\begin{equation}
\begin{array}{rccccccl}
A\!&\!A&\!\emptyset & \to & A\!&\!\emptyset\!&\!A &
\mbox{with rate } r \, , \\
\emptyset\!&\!A\!&\!\emptyset & \to & \emptyset\!&\!\emptyset\!&\!A &
 \mbox{with rate } q \, , \\
\emptyset\!&\!\emptyset\!&\!A & \to & \emptyset\!&\!A\!&\!\emptyset &
\mbox{with rate } p
\, .
\end{array}
\end{equation}
By identifying vacancies with up-spins and particles with
down-spins, these dynamics may be interpreted as a non-equilibrium
spin-relaxation process. The choice $p=0$ is a special case of the
kinetic Ising models of Ref. \cite{Katz84}, with $r=q=1/2$ corresponding to
the totally asymmetric
exclusion process (TASEP) \cite{Schm96}. In yet another mapping
one obtains a growth model for a one-dimensional interface (see below).

\begin{figure}[htbp]
\setlength{\unitlength}{0.5mm}
\begin{center}
\begin{picture}(70,65)

\put(5,5){\line(1,0){60.}}
\put(5,5){\line(0,1){60.}}
\put(5,65){\line(1,-1){60.}}

\put(0,0){0}
\put(0,34){$r$}
\put(0,63){1}
\put(34,0){$q$}
\put(64,0){1}

\put (5,35){\circle*{.5}}
\put (6,32.4){\circle*{.5}}
\put (7,30.1){\circle*{.5}}
\put (8,28.0){\circle*{.5}}
\put (9,26.1){\circle*{.5}}
\put (10,24.4){\circle*{.5}}
\put (11,22.7){\circle*{.5}}
\put (12.5,20.6){\circle*{.5}}
\put (14,18.6){\circle*{.5}}
\put (15.5,16.9){\circle*{.5}}
\put (17,15.4){\circle*{.5}}
\put (18.5,14.0){\circle*{.5}}
\put (20,12.7){\circle*{.5}}
\put (21.5,11.6){\circle*{.5}}
\put (23,10.6){\circle*{.5}}
\put (24.5,9.7){\circle*{.5}}
\put (26,8.8){\circle*{.5}}
\put (27.5,8.1){\circle*{.5}}
\put (29,7.3){\circle*{.5}}
\put (30.5,6.7){\circle*{.5}}
\put (32,6.0){\circle*{.5}}
\put (35,5){\circle*{.5}}

\put(10,10){{\em I}}
\put(25,25){{\em II}}

\end{picture}
\end{center}
\caption{Schematic phase diagram with second order phase transition line
(dotted curve) between
the disordered phase {\em II} and phase
{\em I} with spontaneously broken $Z_2$-symmetry and non-vanishing order
parameter $\Delta$ (cf. Eq.~(\protect\ref{1})). At $r=0$, there is a
transition at $q=1/2$ to phase
separation with two ferromagnetically ordered areas and
spontaneous breaking of translational invariance.
\label{phase}}
\end{figure}
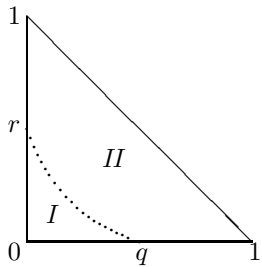

Our interest is in the stationary behavior of the half-filled system, i.e. the
asymptotic state of the system reached at very large times.
A thorough survey of the phase diagram yields as main
features a phase $(I)$ with spontaneously broken $Z_2$-symmetry between
two antiferromagnetic stationary states and a disordered phase $(II)$
(Fig.~\ref{phase}). As we shall argue below, the transition
line separating the two phases belongs to the universality class
of directed percolation except for $r=0$, where the
universality class is different.

More specifically, we found that
the stationary state can be calculated
exactly along the four lines $r=0$, $q=0$, $q=1/2$ and $p=0$
(Fig.~\ref{phase}).
(i) For $q=1/2$, the system is disordered, and the stationary states are
uncorrelated
product measures. (ii) For $p=0$, the stationary distribution is that of a
one-dimensional Ising model \cite{Katz84}. Correlations are short-ranged with
divergent correlation lengths only at the extremal points $q=1$
(phase separation into regimes with complete ferromagnetic order but opposite
magnetization) and $r=1$ (complete antiferromagnetic order), respectively.
(iii) The stationary state along $q=0$ is also
antiferromagnetically ordered, but at $r=p=1/2$ there is an interesting phase
transition in the dynamics of the system. Evidently, for small $q$, transitions
between the two antiferromagnetic states $A\emptyset A\emptyset A\emptyset
\dots$ and $\emptyset A\emptyset A\emptyset A \dots$ are possible
with finite probability, if the system is {\em finite}. However, for $r > 1/2$
the flipping time between these two states diverges with a power law in
system size $L$, whereas for $r < 1/2$ this flipping time diverges
exponentially
in system size. This is a signature for spontaneous symmetry breaking
(and associated ergodicity breaking in the thermodynamic limit), even
away from the line $q=0$. (iv) Along $r=0$ the minimal height of the
corresponding growth model is conserved \cite{Kodu98}. As in the related class
of models of Ref. \cite{Kodu98}, the dynamics satisfies
detailed balance with respect to an energy functional
which is proportional to the area under
the interface. The point $p=q=1/2$
(corresponding to a change in the sign of the energy $E$)
marks the transition from an antiferromagnetic state
to a state where complete phase ordering takes place
and translational invariance is spontaneously broken. This transition is
analogous to the wetting transition of Ref.~\cite{Hinr97}.

This summary of exact results demonstrates the rich behavior that even
rather simple homogeneous lattice gases may show and also indicates a
certain degree of universality of these
phenomena in 1D non-equilibrium systems. Here, we want to
discuss the behavior of the system as it crosses the phase transition line
between the broken symmetry phase {\em I} and the disordered phase {\em II}.
We shall focus on the line $r=q$ with the limiting cases $r=q=1/2$
(usual right hopping TASEP with uncorrelated disordered stationary state) and
$r=q=0$ (left hopping TASEP with next-nearest-neighbour repulsion and
fully ordered stationary states).
We performed Monte-Carlo simulations for half-filled
periodic systems of size $L=2^n$,
mostly with $n=10$.
Expectation values were averaged over $4000L$ rounds after a
transient period of at least the same duration.

We study the quantity
\begin{equation}
 \Delta(t) = \frac{1}{t} \int_0^t \!\! dt \, \frac{2}{L}
 \sum_{i=1}^L (-1)^{i} \langle \, n_i(t) \, \rangle .
\end{equation}
where $n_i=0$ corresponds to an empty site $i$ and $n_i=1$ to an
occupied one.
In the limit $t \to \infty$, it corresponds to the non-conserved
order parameter $2/L \sum_i (-1)^i \langle \, n_i \, \rangle$,
which is the stationary difference in sublattice particle densities
(the `staggered magnetization' in spin language).
Because of ergodicity, the stationary value of the order parameter in
a {\em finite} system vanishes by symmetry. However,
as a signature of spontaneously broken
symmetry in the thermodynamic limit, one expects an initial decay to
some quasi-stationary value $\Delta_0$,
before $\Delta$ eventually approaches zero
for very long times (exponentially large in system size).
On the other hand, in the disordered phase one expects an initially
ordered state with $\Delta = 1$ to rapidly disorder, i.e. one expects
$\Delta$ to decay quickly to zero.

A second quantity of interest is the stationary particle current which,
according to the definition (1) of the process, on the line $r=q=(1-p)/2$
is given by $j(q) = q \langle \, n_i(1-n_{i+1}) \,
\rangle - (1-2q)\langle \, (1-n_{i-1})(1-n_{i}) n_{i+1}\, \rangle$.
Clearly, $j(0) = 0$ and $j(1/2) = 1/8$, up to a
small finite-size correction of order $1/L$. The presence of spontaneous
symmetry breaking suggests $j = 0$ for all $q \leq q_c$ (up to exponentially
small corrections in system size), since any finite current would lead to a
transition between the two degenerate stationary states with
$\Delta = \pm \Delta_0$ within a finite time.

\begin{figure}[htbp]
\setlength{\unitlength}{8.2mm}
\hspace*{-0.5\unitlength}
\epsfig{width=3.8\unitlength, angle=-90,
      file=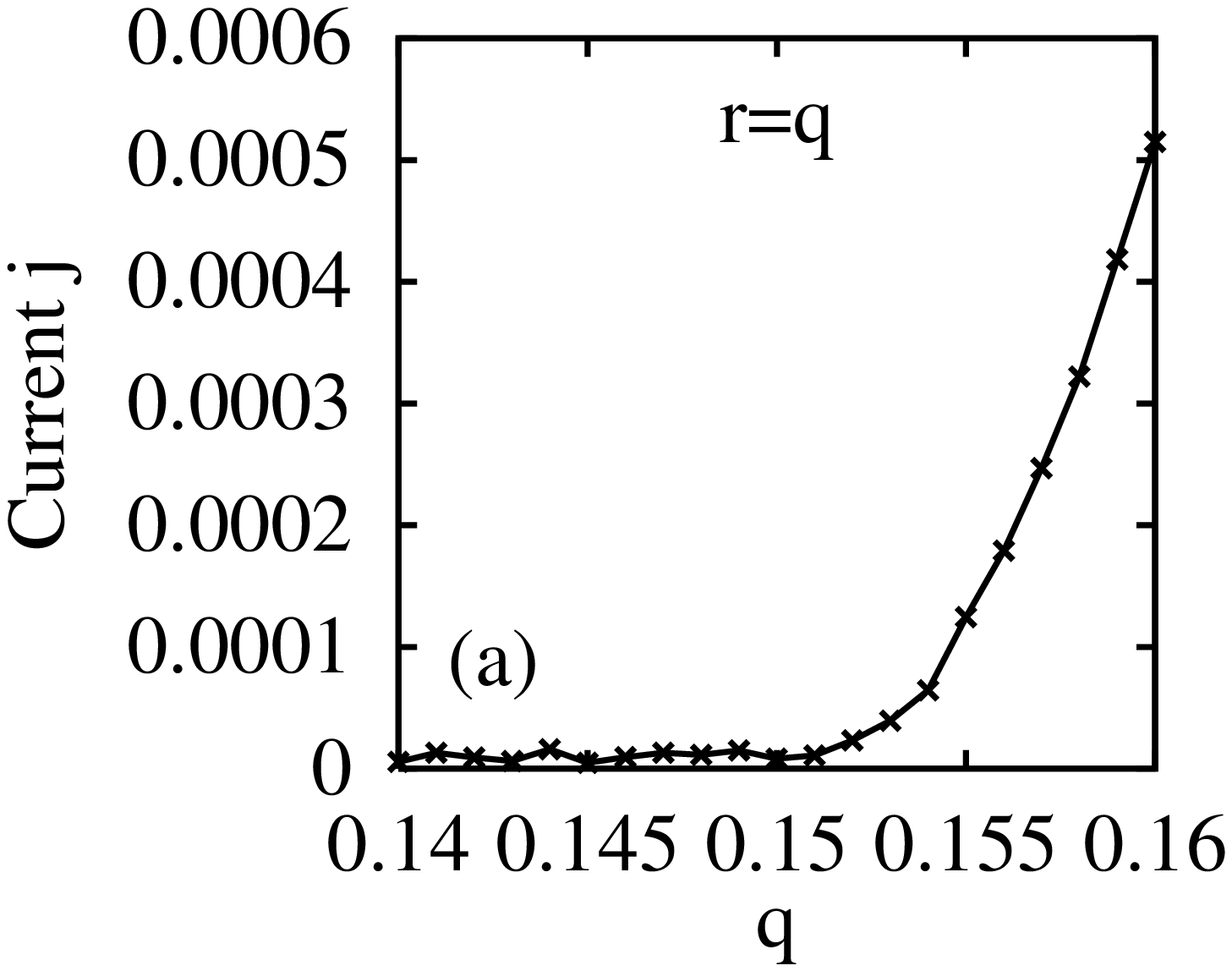}
\hspace*{0.5\unitlength}
\epsfig{width=3.8\unitlength, angle=-90,
      file=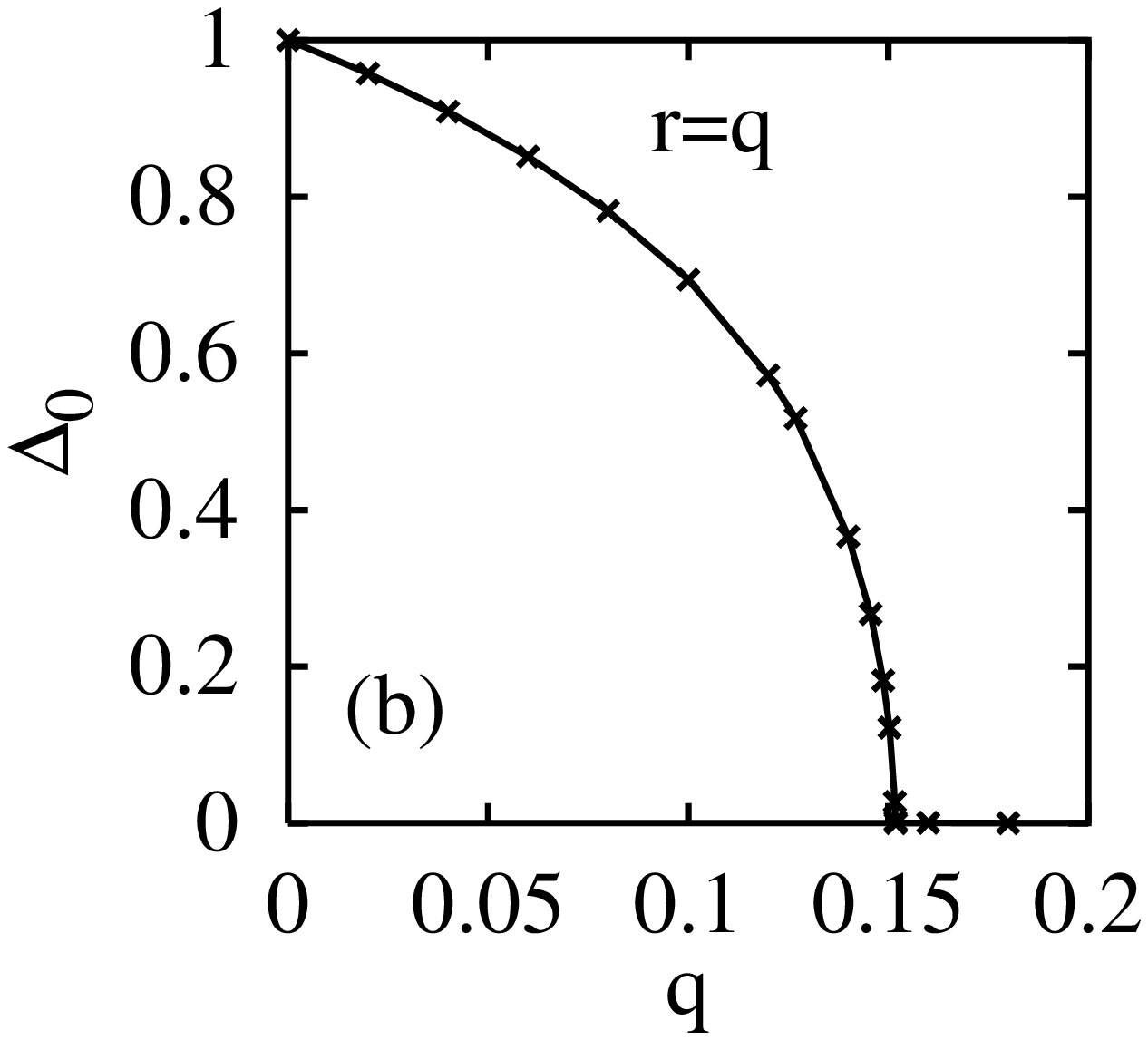}
\hspace*{0.5\unitlength}
\epsfig{height=5.5\unitlength, angle=-90,
      bbllx=120pt, bblly=50pt, bburx=554pt, bbury=700pt,
      file=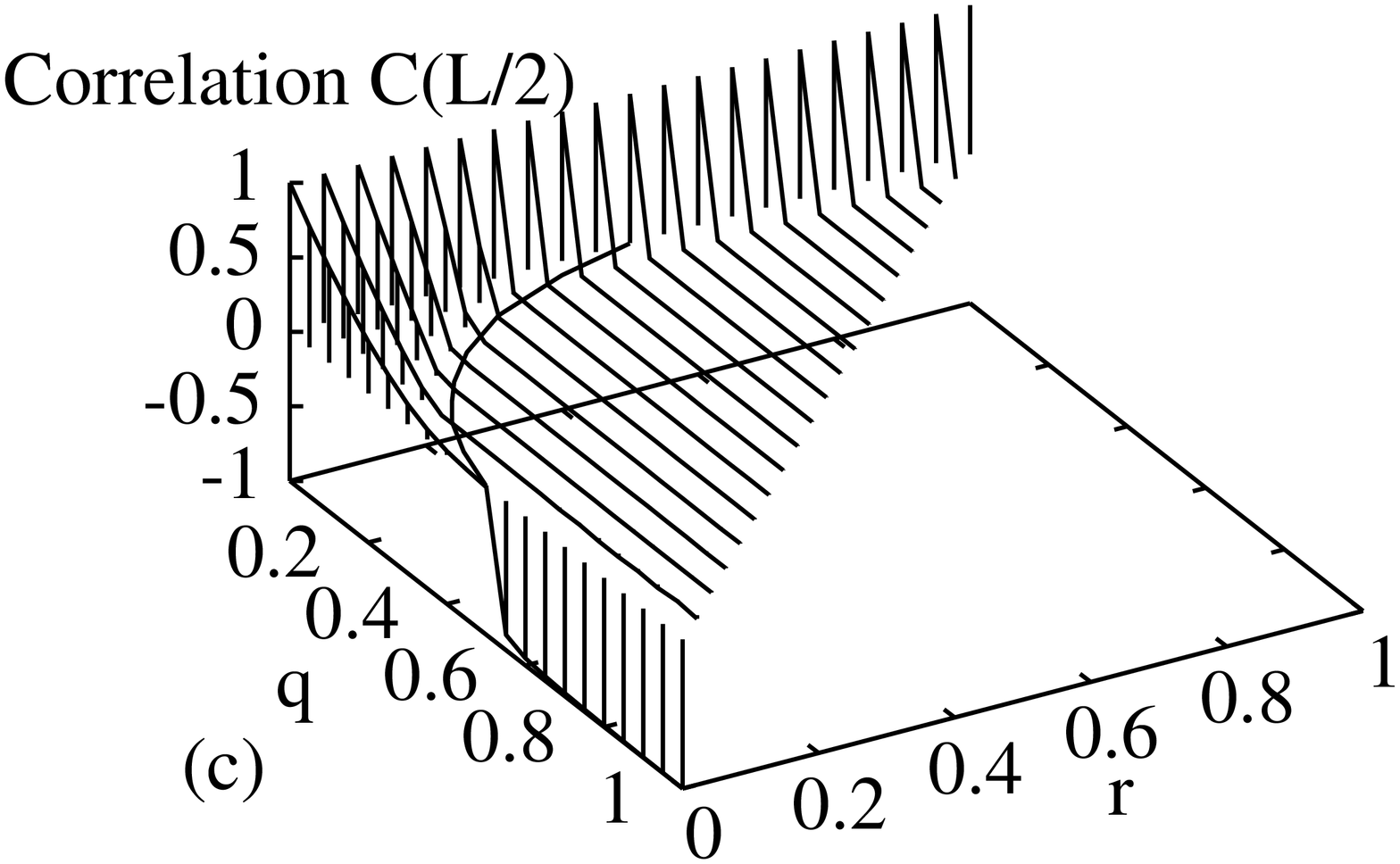}
\epsfig{height=4.2\unitlength, angle=-90, 
      file=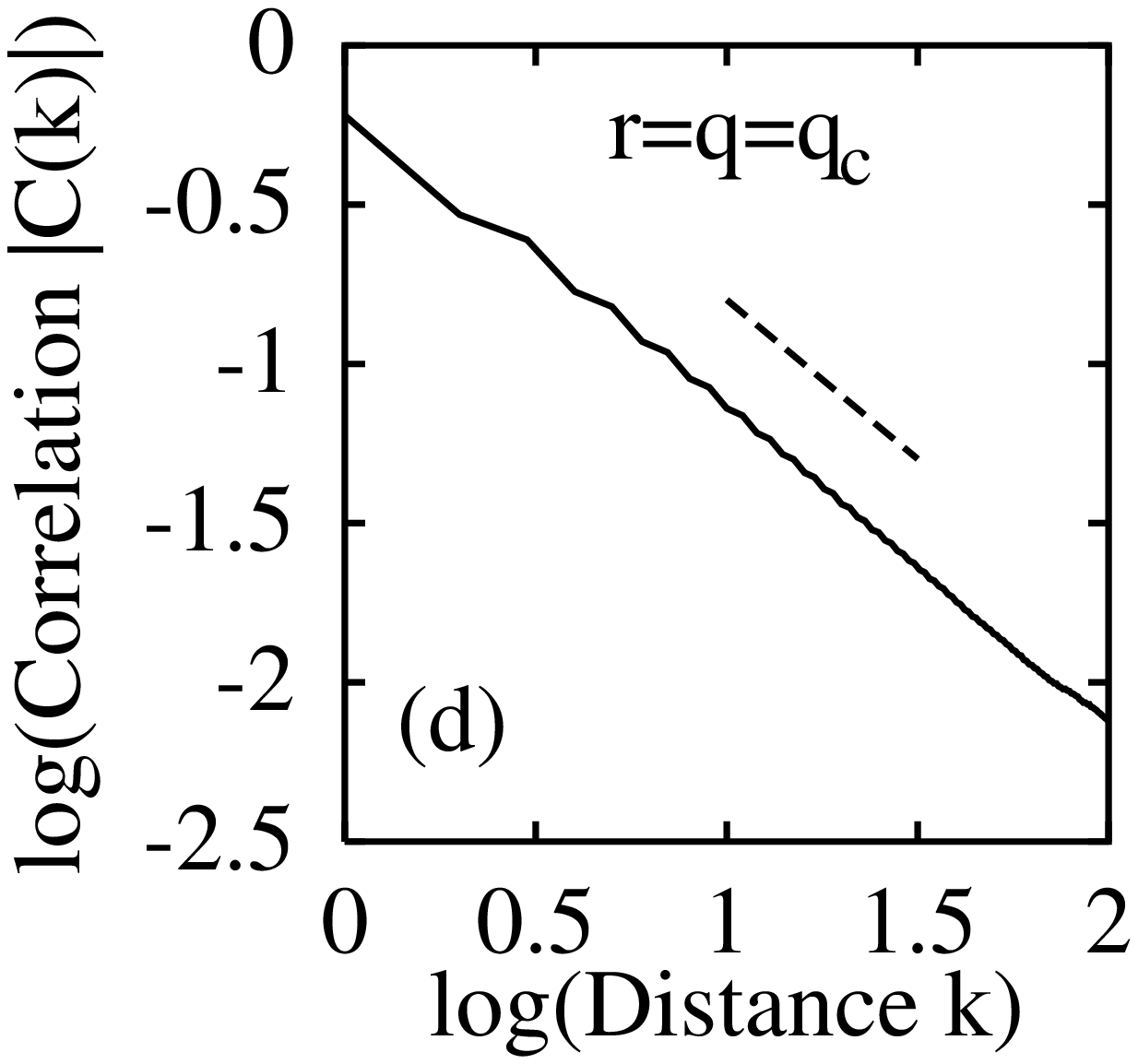}
\caption[]{(a) Stationary current and (b) order parameter
along the line $r=q$. Below: Correlation function $C(k)$ (c) for
maximal distance $k=L/2$ as a function of $r$ and $q$ and (d)
as a function of $k$ for $r=q=q_c$. The dashed line corresponds to a
slope of $-1$. The measured data are connected by straight lines as a guide
for the eye.
\label{MC}}
\end{figure}

This intuitive picture is well-supported by our Monte Carlo simulations
(Fig.~\ref{MC}). The current $j$ vanishes in phase {\em I} and the order
parameter $\Delta_0$ vanishes in phase {\em II}. We find a phase
transition point $q_c = 0.1515\pm 0.0005$ for $r=q$, above which
the current decays with a power law
\begin{equation}
j \sim (q-q_c)^y
\label{2}
\end{equation}
with $y \approx 1.7\pm 0.1$. Approaching the critical point $q_c$ from below,
$\Delta_0$ decays with a power law
\begin{equation}
\Delta_0 \sim (q_c-q)^\theta
\label{1}
\end{equation}
with $\theta \approx 0.54\pm 0.04$.
To investigate whether this continuous bulk phase transition is
accompanied by spatial long-range order---as one would
expect in an equilibrium system---we examine the stationary density correlation
function $C(k) = 4( \langle \, n_i n_{i+k} \, \rangle - \langle \, n_i  \,
\rangle\langle \, n_{i+k} \, \rangle)$ which turns out to decay to a
non-zero value below $q_c$. At the critical point, correlations decay
algebraically
\begin{equation}
C(k) \sim k^{-\gamma} \, ,
\label{3}
\end{equation}
where $\gamma \approx 1.0 \pm 0.1$ (Fig.~ \ref{MC})
\cite{error}.

\begin{figure}[htbp]
\setlength{\unitlength}{0.5mm}
\begin{center}
\begin{picture}(100,50)(-5,5)
\multiput (10,9)(10,0){9}{\line(0,1){2.}}
\put (10,10){\line(1,0){8.}}
\put (15,10){\circle*{2}}
\put (16,10){\line(1,0){8.}}
\put (25,10){\circle{2}}
\put (26,10){\line(1,0){8.}}
\put (35,10){\circle{2}}
\put (36,10){\line(1,0){8.}}
\put (45,10){\circle*{2}}
\put (46,10){\line(1,0){8.}}
\put (55,10){\circle{2}}
\put (56,10){\line(1,0){8.}}
\put (65,10){\circle*{2}}
\put (66,10){\line(1,0){8.}}
\put (75,10){\circle*{2}}
\put (76,10){\line(1,0){8.}}
\put (85,10){\circle{2}}
\put (86,10){\line(1,0){4.}}

\multiput (10,12.5)(0.0,5.0){6}{\line(0,1){2.5}}
\multiput (20,12.5)(0.0,5.0){4}{\line(0,1){2.5}}
\multiput (30,12.5)(0.0,5.0){6}{\line(0,1){2.5}}
\multiput (40,12.5)(0.0,5.0){8}{\line(0,1){2.5}}
\multiput (50,12.5)(0.0,5.0){6}{\line(0,1){2.5}}
\multiput (60,12.5)(0.0,5.0){4}{\line(0,1){2.5}}
\multiput (70,12.5)(0.0,5.0){6}{\line(0,1){2.5}}
\multiput (80,12.5)(0.0,5.0){4}{\line(0,1){2.5}}
\multiput (90,12.5)(0.0,5.0){6}{\line(0,1){2.5}}
\thicklines
\put (18,6){\vector(1,0){4.0}}
\put (62,6){\vector(-1,0){4.0}}
\put (20,40){\vector(0,1){4.0}}
\put (60,40){\vector(0,-1){4.0}}
\put (10,40){\line(1,-1){10.}}
\put (20,30){\line(1,1){20.}}
\put (40,50){\line(1,-1){10.}}
\put (50,40){\line(1,1){10.}}
\put (60,50){\line(1,-1){20.}}
\put (80,30){\line(1,1){10.}}
\multiput (10,40)(1,1){10}{\circle*{0.1}}
\multiput (20,50)(1,-1){10}{\circle*{0.1}}
\multiput (50,40)(1,-1){10}{\circle*{0.1}}
\multiput (60,30)(1,1){10}{\circle*{0.1}}
\end{picture}
\end{center}
\caption{Mapping between lattice
gas dynamics and interface growth in 1+1 dimensions.
A positive (negative) unit slope belongs to a vacancy
(particle). The interface flips (vertical arrows) correspond to particles
hopping on the lattice (horizontal arrows).\label{int}}
\end{figure}
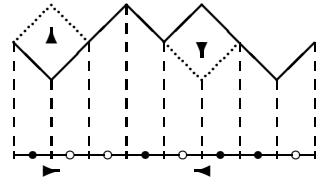

We can gain further insight by considering the mapping to an
interface model \cite{Meak86} which is described by
height difference variables $1-2n_i$ and an additional stochastic variable $h$,
representing the
absolute height of the interface at some reference point. Each time a particle
hops to the right, the local height
increases by two units (deposition), whereas hopping to the left describes
a height decrease (evaporation) (Fig.~\ref{int}). Thus the current
gives
the stationary growth velocity of the interface, while the density correlation
function measures height-gradient
correlations. Growth occurs at local minima with rate $q$,
independently of
the precise nature of the immediate environment. However,
evaporation of particles
does not occur from a ``flat'' part of the interface: The corresponding
process $A  \emptyset A \emptyset  \, \to \,  A A
\emptyset \emptyset$ is forbidden. On a coarse-grained scale this means
that in a locally
flat piece of the interface evaporation is not strong enough to create
little craters which could then further grow. A similar situation (with
different microscopic dynamics) was investigated by
Alon et al. \cite{Alon96}, who found a phase transition between a smooth
phase where no current flows and a rough, growing phase in the universality
class of the KPZ-equation. The transition
is related to directed percolation in 1+1 dimensions and is accompanied by
spontaneous symmetry breaking in the height variable $h$.

\begin{figure}[htbp]
\setlength{\unitlength}{9mm}
\epsfig{width=3.8\unitlength, angle=-90,
      file=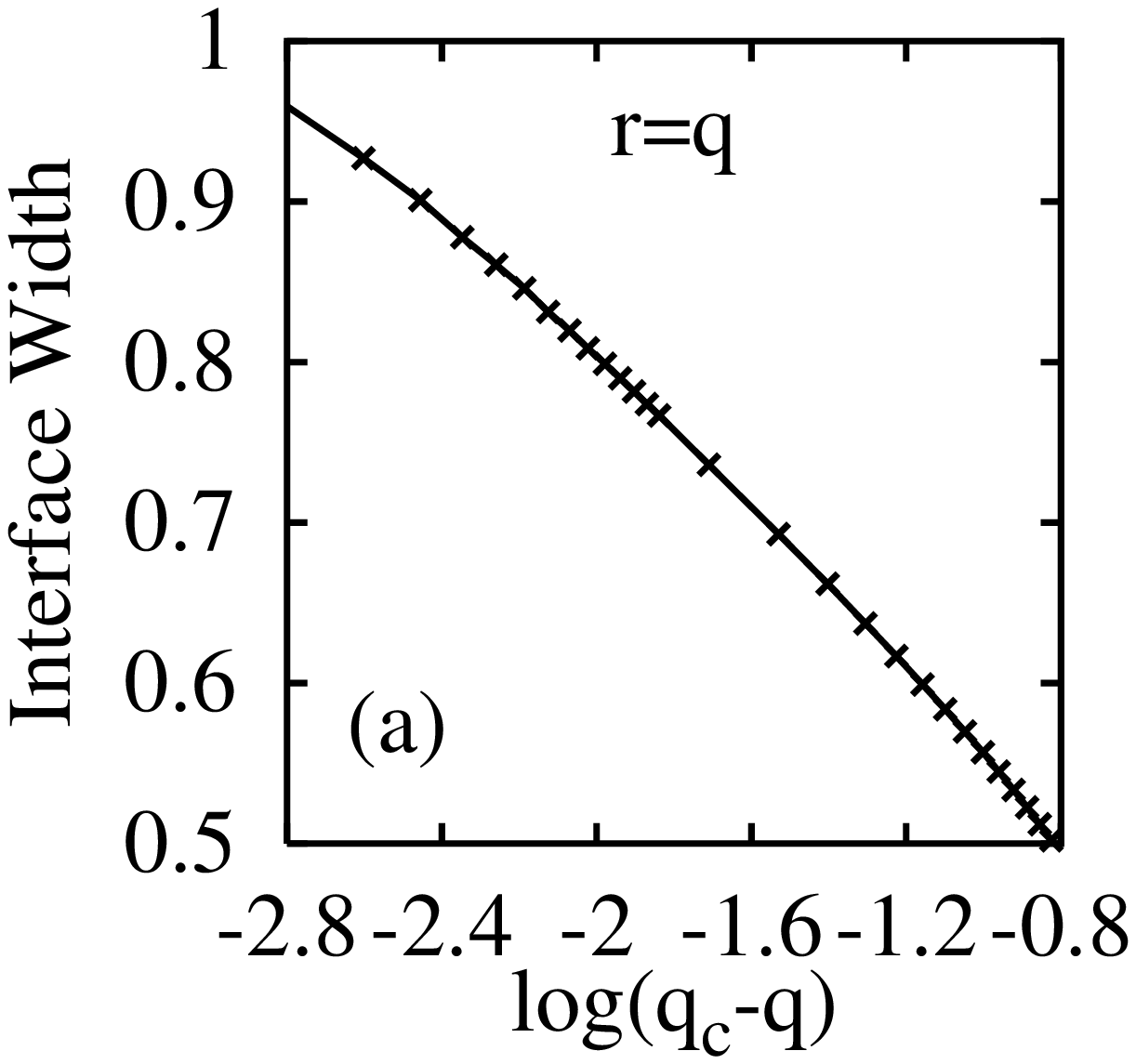}
\epsfig{width=3.8\unitlength, angle=-90,
      file=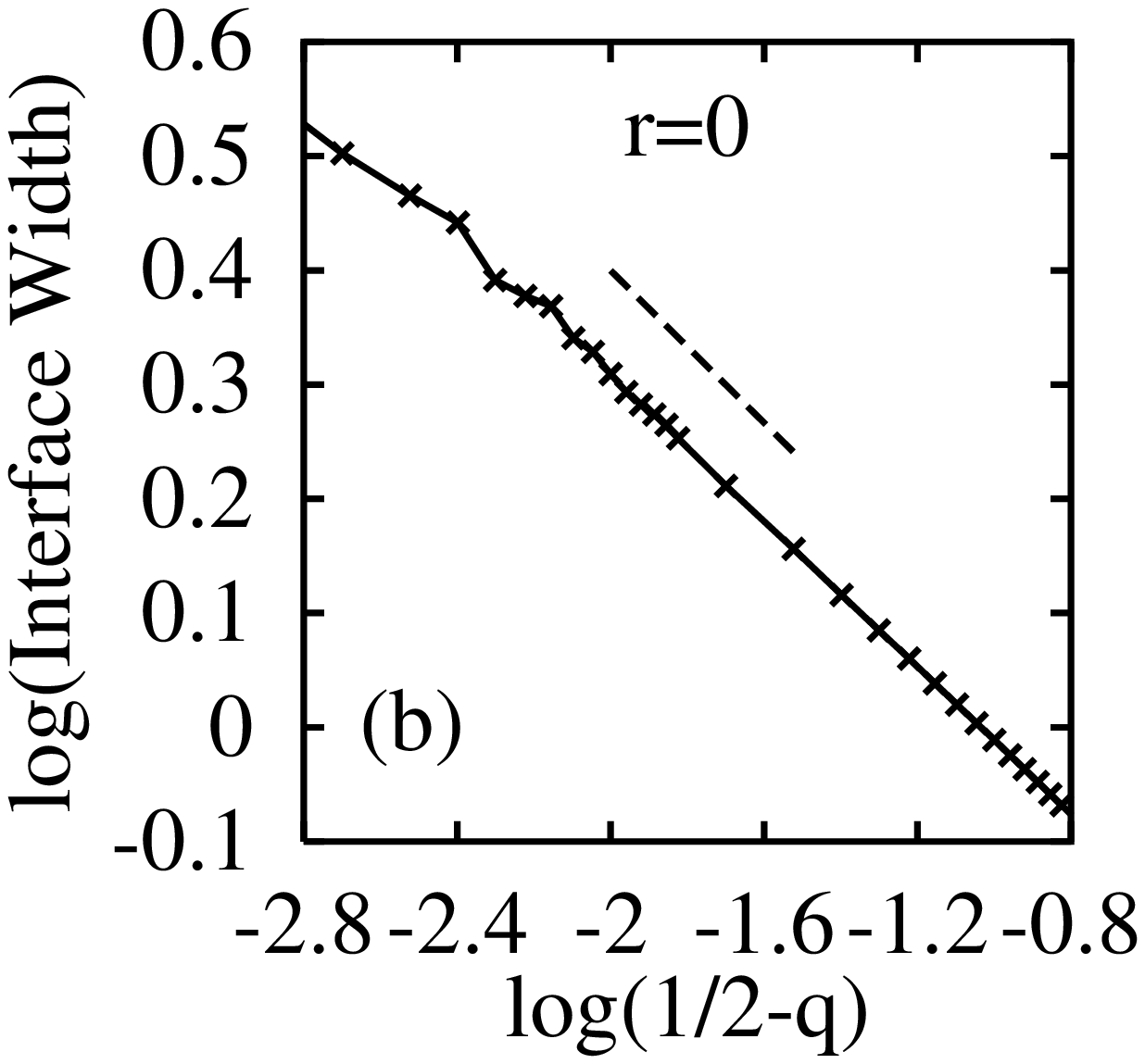}
\caption[]{Interface width close to the phase transition points along
(a) $r=q$, suggesting a logarithmic-like divergence, 
and (b) $r=0$, respectively. The slope of the dashed line is
$-1/3$.\label{width}}
\end{figure}

Here we find similar behavior which is most transparent in the
two limiting cases $q=0$ and $q=1/2$, respectively. The limit $q \to 1/2$
corresponds to the TASEP (growing, rough
interface), which indeed describes interface growth in the KPZ universality
class \cite{Krug91b}.
In the limit $q \to 0$, there is no current and one has spontaneous symmetry
breaking between (macroscopically) flat interfaces on an even or
odd height level, respectively.  We stress, however, that
spontaneous symmetry breaking occurs already on the level of the particle
description, i.e. without reference to the extra height variable.
Assuming universality, one expects \cite{Alon96} the exponent $y$ to be given
by the critical exponent $\nu_{\parallel}\approx 1.73$ of the DP-coherence time
\cite{Kinz83} and also a logarithmic divergence of the interface width
$w = [L^{-1}\sum_i(h_i - L^{-1}\sum_ih_i)^2]^{1/2}$.
This is in agreement
with our results in Eq.~(\ref{2}) and Fig.~\ref{width}.
Also the value (\ref{1}) of the order parameter exponent $\theta$ is
consistent with the result $\theta = 0.55\pm 0.05$ reported in Ref.
\cite{Alon96}, thus independently confirming universality.
Results on the correlation exponent $\gamma$
have not been reported in earlier work.

The transition at $r=0$ is of a different nature.
Here the model satisfies detailed balance and the current vanishes
both above and below the transition. At the phase transition point
$q_c=1/2$ the lattice gas is uncorrelated. Using the interface representation
of the model one can show \cite{Hinr97} that the interface width diverges
algebraically with an exponent $1/3$ as $q$ approaches $1/2$ from below
(Fig.~\ref{width}).

The understanding of $\theta$ and of the new correlation exponent $\gamma$
(which have no conventional interpretation within the framework of directed
percolation), and the behavior of these two quantities at the transition at
$r=0$ have to be addressed in future work. Also the behavior of the system
away from half-filling, where preliminary results suggest the disappearance of
phase {\em I}, is an open issue. Returning to our original
question we conclude at this point that the stationary states of homogeneous
one-dimensional lattice
gas models may exhibit continuous bulk phase transitions with an algebraic
decay of correlations even if interactions are short-ranged. In our
model, this transition results from dynamical constraints
which---unlike in the KLS models---lead to a competition between a
disordering dynamics (the right-hopping process) and processes forcing
the system into either of two antiferromagnetically ordered states
(the restricted left hopping process).
For sufficiently strong ordering
processes, the stationary current ceases to flow and spontaneous symmetry
breaking sets in.

It is interesting to consider yet another mapping of our model,
obtained by mapping particles into vacancies and vice versa on one (either
even or odd) sublattice. The resulting dynamics are that of a new class of
parity-conserving (PC) branching-annihilation
processes $\emptyset AA \rightleftharpoons \emptyset\emptyset\emptyset$ and
$A\emptyset\emptyset  \rightleftharpoons AAA$ with no absorbing state.
In addition to particle-parity conservation (particle number
modulo 2), there is a $U(1)$ symmetry which
results from the particle number conservation of the original hopping process.
Generically, one expects parity-conserving
branching-annihilation processes not to be in the DP universality class, but in
a distinct PC universality class \cite{Card98}. From our results it appears
that, in the presence of additional symmetries, 
the picture of phase transitions
in 1D branching-annihilation processes is more complicated.

We thank M. R. Evans for helpful discussions.
G.M.S. and D.H. thank the Weizmann Institute for kind hospitality.
We are grateful for financial support by the Einstein
Center (G.M.S.), the DFG's Heisenberg program (D.H., grant no.
He 2789/1-1) and support of the Israeli Science Foundation
and the Israel Ministry of Science.


\begin{references}
\bibitem{error}
The errors for the exponents are estimated from
the slopes of $j$, $\Delta_0$, and $C(k)$, respectively,
in double-logarithmic scale
by using the two values $q_c = 0.152$ and
$q_c = 0.151$, respectively, 
for the numerically determined critical point, and using
data for the largest available system size $L=1024$. Finite-size effects
make a more precise determination of the critical point difficult.
\end{references}
\end{document}